\documentclass[a4paper,11pt]{article}
\pdfoutput=1 % if your are submitting a pdflatex (i.e. if you have
\usepackage{jinstpub} % for details on the use of the package, please
\usepackage{amsmath}

\usepackage{float}

\title{\boldmath Aging Study on Resistive Plate Chambers of the CMS Muon Detector for HL-LHC}

\author[n,1]{R. Aly,\note{Corresponding author.}}
\author[n]{,A. Gelmi}
\author[l]{,P. Kumari}
\author[a]{,N. Zaganidis}
\author[a]{A. Samalan}
\author[a]{,M. Tytgat}
\author[b]{,G.A. Alves}
\author[b]{,F. Marujo}
\author[c]{,F. Torres Da Silva De Araujo}
\author[c]{,E.M. Da Costa}
% ...

\affiliation[a]{Ghent University, Dept. of Physics and Astronomy, Proeftuinstraat 86, B-9000 Ghent, Belgium}
\affiliation[b]{Centro Brasileiro Pesquisas Fisicas, R. Dr. Xavier Sigaud, 150 - Urca, Rio de Janeiro - RJ, 22290-180, Brazil}
\affiliation[c]{Dep. de Fisica Nuclear e Altas Energias, Instituto de Fisica, Universidade do Estado do Rio de Janeiro, Rua Sao Francisco Xavier, 524, BR - Rio de Janeiro 20559-900, RJ, Brazil}
% ...
\affiliation[l]{Department of Physics, Panjab University, Chandigarh 160 014, India}
% ...
\affiliation[n]{INFN, Sezione di Bari, Via Orabona 4, IT-70126 Bari, Italy.}
% ...

\emailAdd{reham.aly@cern.ch}

\abstract{In the High Luminosity Large Hadron Collider (HL-LHC) program, during the next years, the instantaneous luminosity will increase up to 5 $\times$ 10$^{34}$ cm$^{-2}$ s$^{-1}$ which means a factor five higher than the nominal LHC luminosity. In that period, the present CMS Resistive Plate Chambers (RPC) system will be subjected to background rates higher than those for which the detectors have been designed, which could affect the detector properties and induce aging effects. To study whether the present RPC system can sustain the hard background conditions during the HL-LHC running period, a dedicated longevity test is ongoing at the CERN Gamma Irradiation Facility, where a few spare RPCs are exposed to high gamma radiation for a long term period to mimic the HL-LHC operational conditions. During the longevity test, the main detector parameters are continuously monitored as a function of the integrated charge. Preliminary results of the study, after having collected a sufficient amount of the expected integrated charge at HL-LHC, will be presented.}


\proceeding{15$^{\text{th}}$ Workshop on Resistive Plate Chambers and Related Detectors RPC2020\\
  10-14 February, 2020\\
  Rome, Italy}

\begin{document}
\maketitle
\flushbottom

\section{Introduction}
\label{sec:intro}

The Muon Tracking System, which lies on the outside of the Compact Muon Solenoid (CMS) experiment \cite{{cms-detector}} at the CERN Large Hadron Collider (LHC), has been designed to provide an efficient muon trigger as well as a precise measurement of muon momentum and charge. It consists of three sub-detectors arranged in barrel and endcap regions: Drift Tubes in the barrel region, Cathode Strip Chambers in the endcap region and 1056 Resistive Plate Chambers (RPC) installed in both regions covering a pseudorapidity region up to $|\eta|$ = 1.9 \cite{cms} \cite{muon-system}.
%A CMS RPC chamber consists of two layers of 2 mm gas gaps with a sheet of segmented copper readout strips sandwiched between them and aligned in the $\eta$ direction. 
A CMS RPC chamber consists of two layers of 2 mm gas gaps with a sheet of segmented copper readout strips sandwiched between them and aligned along the direction parallel to the magnetic field lines. Each gas gap is made of two sheets of High-Pressure-Laminate (HPL) electrodes with 2 mm thickness and filled with a non-flammable three-component gas mixture of 95.2\% freon (C${_2}$H${_2}$F${_4}$), 4.5\% isobutane (i-C${_4}$H$_{10}$), and 0.3\% sulphur hexafluoride (SF${_6}$) with a relative humidity of $\approx$ 40$\%$ \cite{cms}. 
The muon system worked efficiently during the LHC Runs I and II data takings at the nominal luminosity of 1 $\times$ 10$^{34}$ cm$^{-2}$ s$^{-1}$ \cite{muon-performanceRunI,muon-performance1,muon-performance2}. 

\section{Aim of longevity study}

Gas detectors can suffer from aging effects when exposed to high radiation for long time which result in a degradation of detector performance appearing as loss in detector efficiency, increase in dark current\footnote{Dark current is the current produced in the chamber when applying high voltage in the absence of background radiation.} and rise in noise rates. The main reason for this detector performance degradation are the chemical processes that happen inside the electron multiplication region where the gas molecular fragments produced inside avalanches may form polymers growing on the electrodes' surface \cite{HF}. The present CMS RPC system has been certified for 10 years of LHC operation at a maximum background rate of 300 Hz/cm$^2$ and a total integrated charge of 50 mC/cm$^2$. Based on the data collected by CMS during LHC Run II and assuming a linear dependence of the background rates as a function of the instantaneous luminosity, the expected background rates and integrated charge at HL-LHC will be about 600  Hz/cm$^2$ and 840 mC/cm$^2$, respectively (including a safety factor of three) \cite{HL-LHC}. 
In those operating conditions a non-recoverable aging effects can appear, due to the higher collision rates and pile-up\footnote{Pile-up known as more than one proton-proton collision takes place in the same beam crossing.}, that can affect the detector performance and properties.
%The operating conditions during the HL-LHC phase will be much harder, in terms of high radiation - background rate - Pile-up - probable aging, with respect to those for which the detectors had been designed and can cause a non-recovable aging effects that can affect the detector performance and properties. 
Therefore, a long term irradiation test has been carried out to study whether the present RPC detectors can survive the hard background conditions during the HL-LHC running period. During the longevity studies, the main detector parameters and performance are monitored as a function of the integrated charge to spot any possible aging effects.

\section{Longevity test setup and procedure}

\begin{figure}[htbp]
\centering % \begin{center}/\end{center} takes some additional vertical space
\includegraphics[width=.5\textwidth,origin=c]{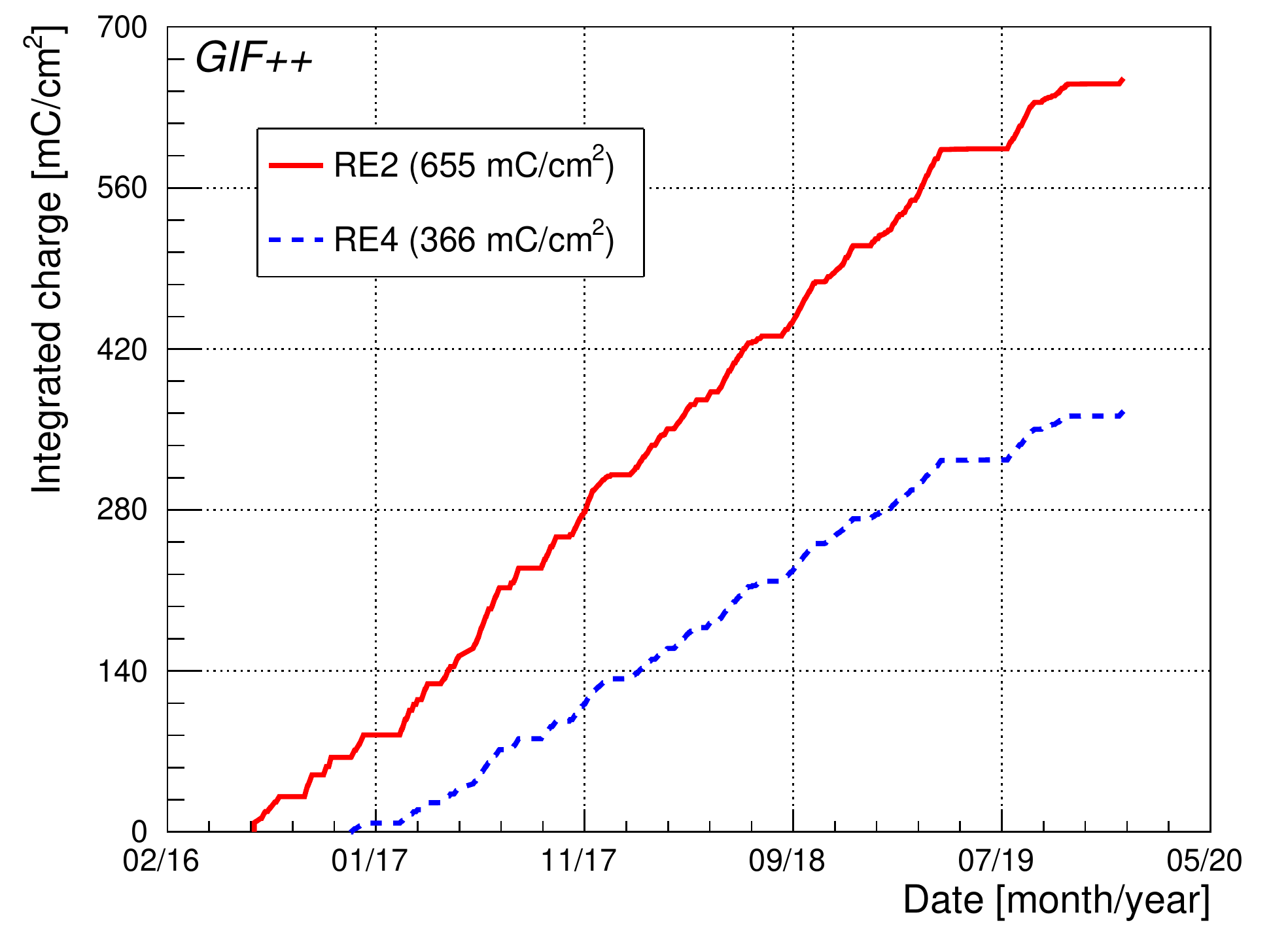}
%\qquad
\setlength\abovecaptionskip{+1.5pt}
\caption{\label{fig:qint} Integrated charge versus time, accumulated during the longevity test at GIF++ for RE2/2 (solid red line) and RE4/2 (dashed blue line) chambers. The RE4/2 chamber has been turned on a few months later because of total gas flow limitations. Different slopes account for different irradiation conditions during data taking.}
\end{figure}

A long term irradiation test has been started at the CERN Gamma Irradiation Facility (GIF++) which allows to test real size detectors. The GIF++ is a unique place equipped by a gamma source (13 TBq Cs-137) and a system of movable filters for varying the gamma flux which allow to test the detectors in a background condition similar to the ones at HL-LHC \cite{gif++}. In addition to the gamma source, a 100 GeV muon beam is provided 3-4 times per year for detector performance studies. The GIF++ facility provides also a controlled monitor of the environmental parameters during irradiation such as  temperature and pressure.
Four spare RPC chambers have been irradiated since July 2016, two RE2/2s and two RE4/2s \cite{cms-detector}\cite{HL-LHC} which are from the endcap region where the maximum background radiation rates are expected. The detectors are trapezoidally shaped with height = 1687 mm , long side = 979 mm and short side = 684 mm.
Two different RPC production types have been tested, since the endcap RPC production was done in two different periods, i.e. all RPCs in endcap were done in 2005 except RE4/2 and RE4/3 which were made later in 2012/2013. Two chambers (one RE2 and one RE4) are continuously under irradiation while the other two chambers of the same type are kept as reference and switched on from time to time.
All the detectors are flushed continuously with gas where the detectors are currently running with gas humidity $\approx$ 60$\%$ and 3 gas volume exchange per hour for irradiated chambers and one gas volume exchange per hour for reference chambers.
The detector parameters  (such as dark current, noise rate, current and count rates at several background conditions) are monitored continuously and compared with the measurements from the reference chambers to spot any degradation in the detector parameters due to long term irradiation. Moreover, when the muon beam at GIF++ is available, the detector performance is studied at different irradiation fluxes.
%also 3-4 times per year during the muon test beam the detector performance studies are performed.
%The detectors are currently running with gas humidity $\approx$ 60$\%$ and 3 gas volume exchange per hour for irradiated chambers and one gas volume exchange per hour for reference chambers. 
Since the gamma flux is uniformly distributed over the detector surface, the integrated charge is calculated as the average density current accumulated in time in the three gaps that constitute the detector. The integrated charge collected from the beginning of irradiation till February 2020 are about 655 and 366 mC/cm$^2$ for RE2 and RE4 chambers respectively as shown in Fig. \ref{fig:qint} that correspond to approximately 78$\%$  and 44$\%$ of the expected integrated charges at HL-LHC. 

\section{Detector parameter monitoring}

\subsection{Dark current and noise rate}

\begin{figure}[htbp]
\centering % \begin{center}/\end{center} takes some additional vertical space
\includegraphics[width=.45\textwidth,origin=c]{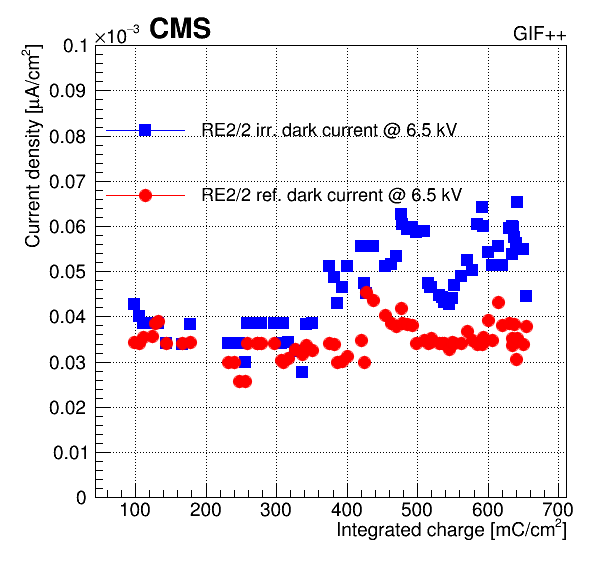}
%\qquad
\includegraphics[width=.45\textwidth,origin=c]{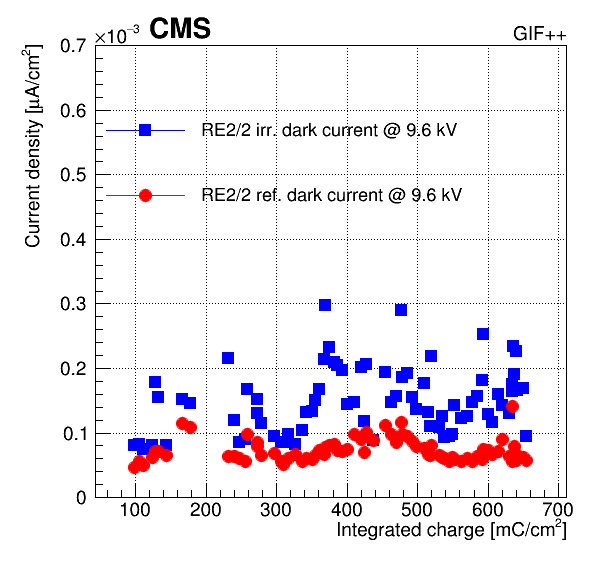}
% "\includegraphics" from the "graphicx" permits to crop (trim+clip)
% and rotate (angle) and image (and much more)
\caption{\label{fig:darkcurrent} Dark current density for RE2 irradiated (blue squares) and reference (red circles) chambers as a function of collected integrated charge at 6.5 kV (left) and at 9.6 kV (right).}
\end{figure}

The dark current and noise rate are monitored periodically in order to spot any aging effect due to irradiation. The dark current density "Current normalized to the surface area" for RE2 both irradiated and reference chambers as a function of collected integrated charge is shown in Fig. \ref{fig:darkcurrent} . The dark currents were measured at 6.5 kV (left), which represent the ohmic contribution, and at 9.6 kV (right), which includes the gas amplification.
 %at 6.5 kV which represents the ohmic contribution, and on the right at 9.6 kV which includes also the gas amplification. 

\begin{figure}[htbp]
\centering % \begin{center}/\end{center} takes some additional vertical space
\includegraphics[width=.45\textwidth,origin=c]{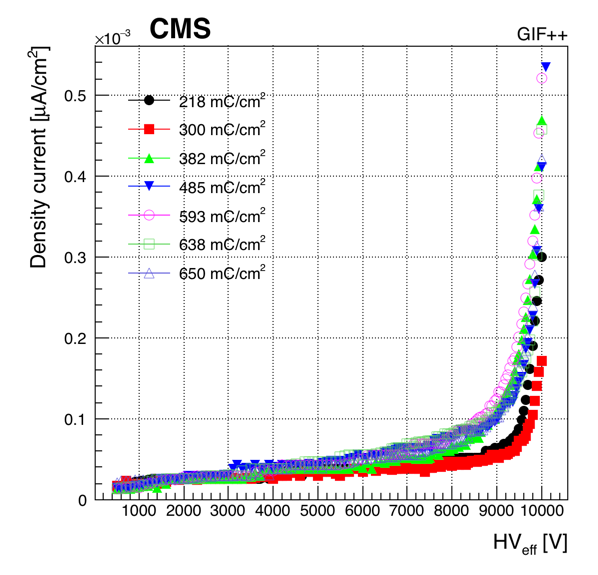}
%\qquad
\includegraphics[width=.45\textwidth,origin=c]{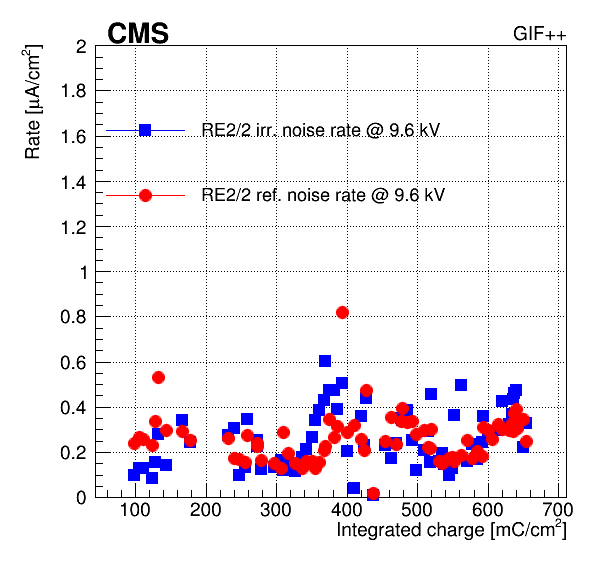}
\caption{\label{fig:hv_scan} Dark current density monitored as a function of the effective high voltage at different values of collected integrated charge for RE2 irradiated chamber (left) and (right) average noise rate as a function of collected integrated charge for RE2 irradiated (blue squares) and reference chambers (red circles). }
\end{figure}

 \noindent The dark current is almost stable in time with small acceptable variations of dark current level since the beginning of irradiation. Figure \ref{fig:hv_scan} (left) shows the dark current density monitored as function of effective high voltage (voltage normalized at the standard temperature  20 $^o$C and pressure 990 $mbar$ \cite{HV_Norm}) at different values of collected integrated charge. The dark current is almost stable with time with small acceptable variations of dark current level since the beginning of irradiation. Figure \ref{fig:hv_scan} (right) shows the average noise rate for RE2 irradiated (blue) and reference (red) chamber as a function of collected integrated charge, the average noise rate is stable with time and less than 1 Hz/cm$^2$.

\subsection{Resistivity and current}

The current with the presence of background radiation is measured periodically as well. In addition, the electrode's resistivity is measured several times per year since it is a crucial parameter that influences the RPC performance. The resistivity is measured by filling up the detector with pure argon and operating in a self-sustaining streamer mode, when the gas quenching components such as the isobutane are removed the streamers propagate all over the detector area and by measuring the current and the applied high voltage the resistance can be calculated and hence the resistivity. 
The measured resistivity values are normalized to 20 $^o$C to allow comparing the values at different temperature conditions \cite{R_Temp}. %To exclude the dependence on the external parameters, the ratios of the irradiated and the reference chambers is measured as a function of the integrated charge.

\begin{figure}[htbp]
\centering % \begin{center}/\end{center} takes some additional vertical space
\includegraphics[width=.5\textwidth,origin=c]{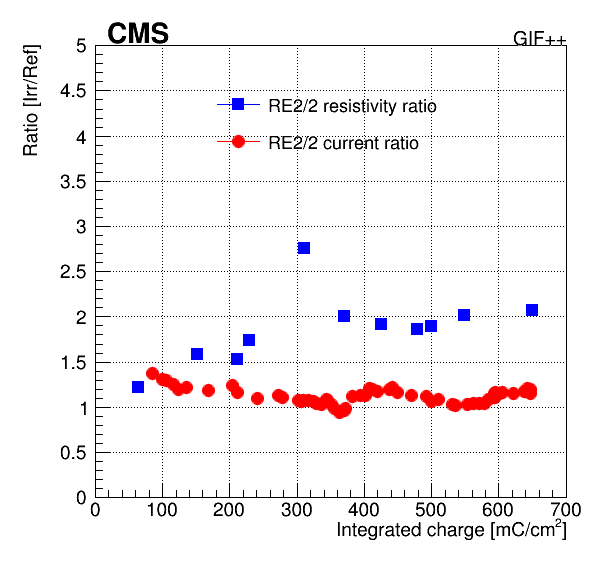}
%\qquad
\caption{\label{fig:resistivity} Resistivity ratio (blue squares) and current ratio (red circles) between RE2 irradiated and reference chambers as a function of collected integrated charge.}
\end{figure}

Figure \ref{fig:resistivity} shows the resistivity ratio and the current ratio " current under gamma background rate of about 600 Hz/cm$^2$"  between irradiated and reference chambers, these ratios are taken to exclude the effect of external parameters.  An increase in the resistivity was observed in the irradiated chamber in the first irradiation period, up to $\approx$ 300 mC/cm$^2$, when the detectors operated in similar conditions as in CMS: one gas volume exchange per hour and $\approx$ 35-45$\%$ of relative gas humidity. These operating conditions were optimized for CMS, but they are not optimal with respect to the high gamma background rate ( $\approx$ 600  Hz/cm$^2$) at GIF++. Therefore, these conditions led to a drying up of the HPL plates with the consequent resistivity increase, which is also confirmed by the decrease of the currents.
At $\approx$ 300 mC/cm$^2$, the relative gas humidity was increased and maintained at $\approx$ 60$\%$, and the gas flow was increased in the irradiated chamber at three gas volume exchanges per hour. The combination of these effects allowed to reduce the HPL resistivity and mitigate the variations, proving that the observed resistivity increase was depending on the operating conditions and it is a recoverable effect.
 
\section{Detector performance monitoring}

\begin{figure}[htbp]
\centering % \begin{center}/\end{center} takes some additional vertical space
\includegraphics[width=.45\textwidth,origin=c]{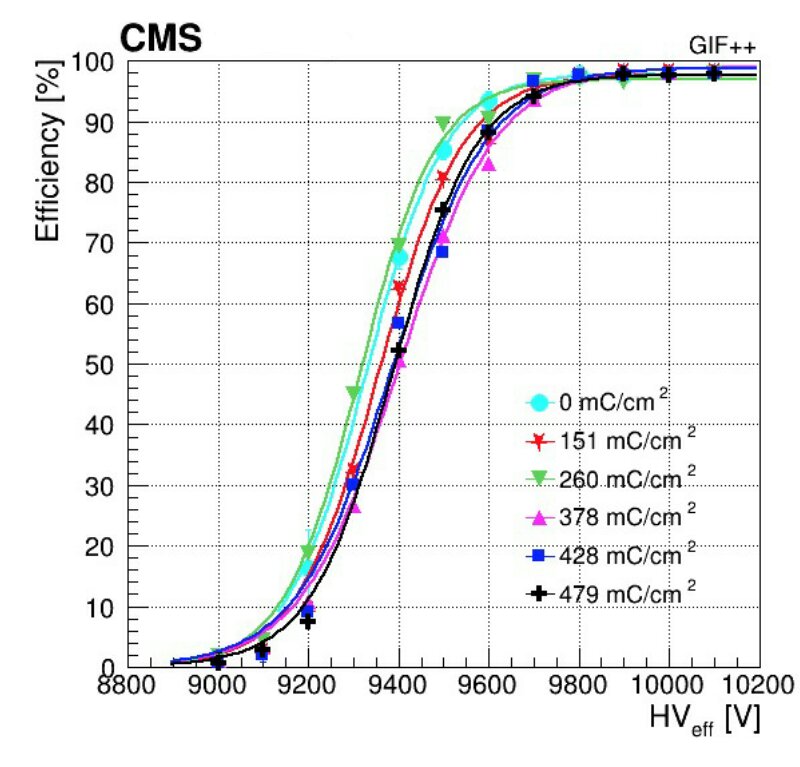}
%\qquad
\includegraphics[width=.45\textwidth,origin=c]{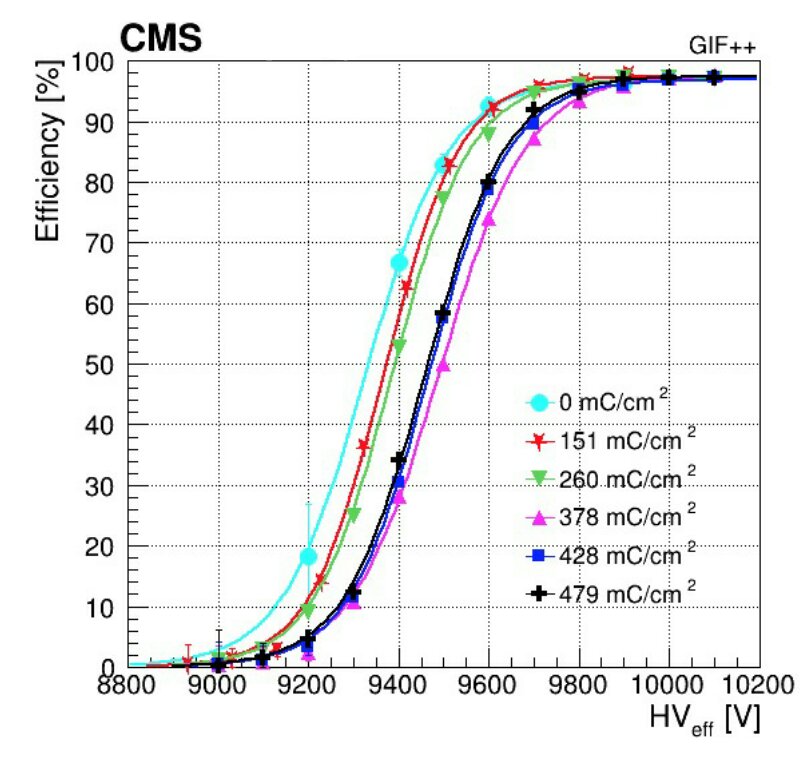}
% "\includegraphics" from the "graphicx" permits to crop (trim+clip)
% and rotate (angle) and image (and much more)
\caption{\label{fig:efficency} RE2/2 irradiated chamber efficiency as a function of the effective HV, with no irradiation (left) and under a gamma background rate of about 600 Hz/cm$^2$ (right) at different irradiation periods. }
\end{figure}

The detector performance has been measured during test beams before irradiation and at different periods of irradiation.  The last measurement was done at 479 mC/cm$^2$ at the last muon beam in GIF++ in 2018. Figure \ref{fig:efficency} shows the RE2 irradiated chamber efficiency measured as a function of the effective high voltage without background radiation (left) and in the presence of 600  Hz/cm$^2$ background (right) at different values of collected integrated charge.  The efficiency is stable in time in the absence of the background radiation and we do not observe any working point shift \cite{HV_Norm2}, while in the case of presence of background, the efficiency is stable at the working point but we observe a working point shift of 100 V after collecting 260 mC/cm$^2$ of integrated charge. The working point shift is related to the resistance (R) of electrodes increase observed at 300 mC/cm$^2$ of integrated charge as shown in Fig. \ref{fig:resistivity}. This increase of R causes an increase of the voltage drop (RI), where I is the total current, on the effective voltage (HV) applied to the electrodes, and the effective voltage on the gas (HV$_{gas}$) is no longer the same \cite{hvgas1,hvgas2}. The HV$_{gas}$ is defined as:

\begin{equation*}
%HV_{gas} = HV - RI 
\text{HV$_{gas}$ = HV - RI }
\end{equation*}
where R is the bakelite resistance and I is the total current.

\begin{figure}[htbp]
\centering % \begin{center}/\end{center} takes some additional vertical space
\includegraphics[width=.45\textwidth,origin=c]{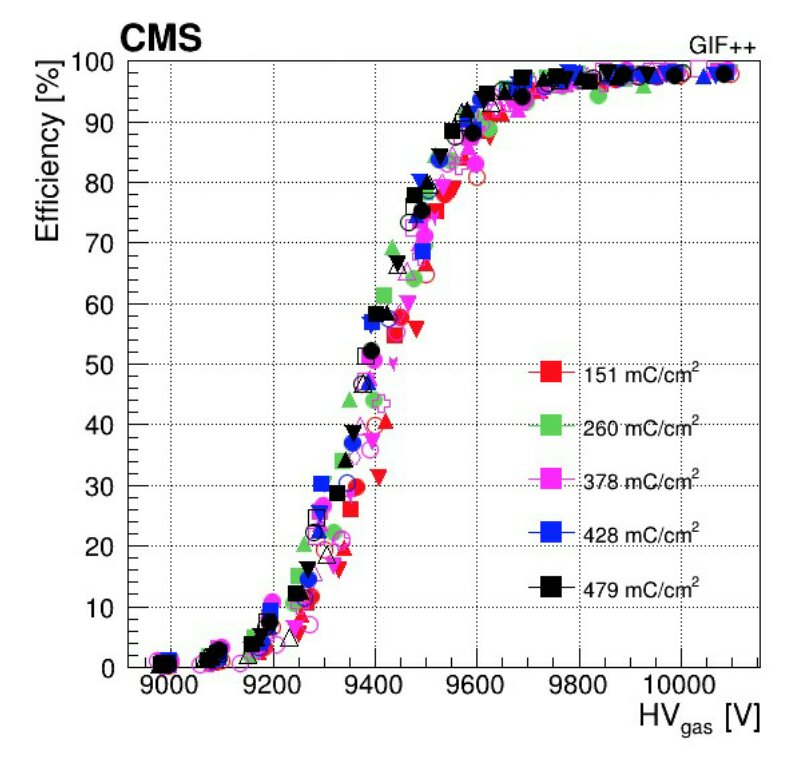}
%\qquad
\includegraphics[width=.45\textwidth,origin=c]{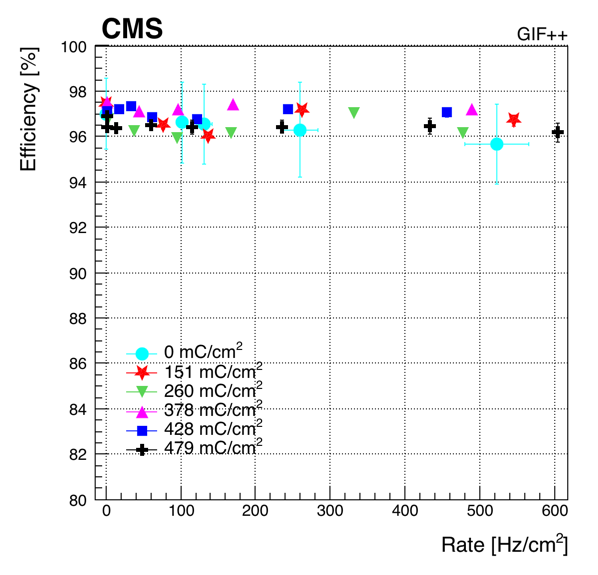}
\caption{\label{fig:hvgas} RE2/2 irradiated chamber efficiency as a function of the HV$_{gas}$ (left)  at different background irradiation rates and different integrated charge values. The RE2 irradiated chamber efficiency at working point as a function of the background rate at different values of collected integrated charge (right) .}
\end{figure}

The detector operation regime is invariant with respect to HV$_{gas}$, therefore the efficiency as a function of HV$_{gas}$ does not depend anymore on the bakelite resistance as shown in Fig. \ref{fig:hvgas} (left) which represents the efficiency at different irradiation periods and different background rates up to 600 Hz/cm$^2$. All the efficiency curves overlap and we do not observe anymore the working point shift, since the R increase effect on the electrodes has been removed.

The RE2 irradiated chamber efficiency at working point is measured at different background rates (up to 600 Hz/cm$^2$) and at different integrated charge values as shown in Fig. \ref{fig:hvgas} (right). The efficiency is stable in time up to the highest background rate expected at HL-LHC (600 Hz/cm$^2$ ).

%The RE2 irradiated chamber efficiency is measured at different background rate and at different time of irradiation periods as shown in figure \ref{fig:eff_vs_rate}. We can see that the efficiency is stable in time up to the highest expected background rate at HL-LHC (600 $Hz/cm^2$ ). 

%\begin{figure}[htbp]
%\centering % \begin{center}/\end{center} takes some additional vertical space
%\includegraphics[width=.4\textwidth,origin=c]{plots/T1_S2_RE2-2_BARC_9_Efficiency_vs_Rate.png}
%\qquad
%\caption{\label{fig:eff_vs_rate} }
%\end{figure}

\section{Conclusion}

Longevity studies on spare resistive plate chambers are ongoing at the CERN Gamma Irradiation Facility under controlled conditions. Preliminary results show no evidence of any aging effect been observed so far. The main detector parameters and performance are stable. The integrated charge collected up to February 2020 represents 78$\%$ of the expected integrated charge at High Luminosity Large Hadron Collider, and more irradiation is needed to complete the study. 

%%\appendix
%%\section{Some title}
%%Please always give a title also for appendices.

\acknowledgments

%This is the most common positions for acknowledgments. A macro is
%available to maintain the same layout and spelling of the heading.

%\paragraph{Note added.} This is also a good position for notes added
%after the paper has been written.

%The measurements leading to these results have been performed at CERN Gamma Irradiation Facility with financial support from AIDA-2020 transnationale access program and INFN funds.

We would like to thank our colleagues from CERN Gamma Irradiation facility where the measurements leading to those results have been performed with financial support from AIDA-2020 transnationale access program and INFN funds. This project has received funding from the European Union's Horizon 2020 Research and Innovation programe under Grant Agreement no. 654168. Also my sincere thanks to all CMS RPC members for their valuable work and RPC2020 organizers for a great conference.


\end{document}